\documentclass[aps,pra,twocolumn,showpacs,showkeywords,superscriptaddress,letterpaper]{revtex4-1}

\usepackage{graphicx}
\usepackage{mdframed}
\usepackage{framed}
\usepackage{indentfirst}
\usepackage{natbib}
\usepackage{amsmath,amssymb}
\usepackage{subfigure}
\usepackage{enumitem}
\usepackage{multirow}
\usepackage{epstopdf}
\begin{document}

\title{Impact of instructional approach on students' epistemologies \\about experimental physics}



\author{Bethany R. Wilcox}
\affiliation{Department of Physics, University of Colorado, 390 UCB, Boulder, CO 80309}

\author{H. J. Lewandowski}
\affiliation{Department of Physics, University of Colorado, 390 UCB, Boulder, CO 80309}
\affiliation{JILA, National Institute of Standards and Technology and University of Colorado, Boulder, CO 80309}

\begin{abstract}
Student learning in undergraduate physics laboratory courses has garnered increased attention within the PER community.  Considerable work has been done to develop curricular materials and pedagogical techniques designed to enhance student learning within laboratory learning environments.  Examples of these transformation efforts include the Investigative Science Learning Environment (ISLE), Modeling Instruction, and integrated lab/lecture environments (e.g., studio physics).  In addition to improving students' understanding of the physics content, lab courses often have an implicit or explicit goal of increasing students' understanding and appreciation of the nature of experimental physics.  We examine students' responses to a laboratory-focused epistemological assessment -- the Colorado Learning Attitudes about Science Survey for Experimental Physics (E-CLASS) -- to explore whether courses using transformed curricula or pedagogy show more expert-like student epistemologies relative to courses using traditional guided labs, as well as how this trend varies based on student major or gender.  Data for this study are drawn from an existing data set of responses to the E-CLASS from multiple courses and institutions. 
\end{abstract}

\maketitle

\section{\label{sec:intro}Introduction}

Introductory laboratory courses represent a unique and important component of the physics curriculum \cite{trumper2003labs} and have been specifically highlighted as critical pieces of the undergraduate curriculum by multiple professional groups \cite{AAPT2015guidelines, olson2012excel, nrc2003bio}.  For many students, these courses provide their first exposure to physics as an experimental science.  As such, these courses can be gateways in terms of both recruiting and retaining students in the physics major.  Moreover, for students who do not continue on to major in physics or another science, introductory laboratory courses often represent one of the only experiences these student will get with the process of scientific experimentation.  Gaining a general understanding of the process and nature of scientific experimentation is an important part of creating citizens who are informed consumers of scientific information. 

For these reasons, introductory physics lab courses often have explicit and/or implicit goals that go beyond just conveying physics content to developing expert-like epistemologies with respect to the nature of experimental physics \cite{trumper2003labs, AAPT2015guidelines}.  However, the traditional guided lab approaches to instruction commonly used in physics lab courses have been heavily critiqued as being cookbook and inauthentic \cite{wieman2015labs,nap2013nrc}.  Inauthentic lab activities can have the unintended side-effect of encouraging epistemologies and expectations about the nature of experimental physics that are inconsistent with those of experts.  Several groups within the PER community have worked to address this issue by designing laboratory learning environments that allow students to engage more authentically in the process of experimental physics.  Examples of these environments include the Investigative Science Learning Environment (ISLE) \cite{etkina2001isle}, Modeling Instruction \cite{wells1995modeling}, and integrated lab/lecture environments such as studio physics \cite{wilson1994studio} or SCALE-UP (Student-Centered Activities for Large Enrollment University Physics) \cite{beichner2000scaleup}.  All of these instructional approaches were either designed with improving students epistemologies about the nature of science as an explicit goal \cite{etkina2001isle, beichner2007scaleup}, or have documented improved shifts in students epistemologies as measured by the CLASS (Colorado Learning Attutides about Science Survey \cite{adams2006class}) or MPEX (Maryland Physics Expectation Survey \cite{redish1998mpex}) \cite{brewe2009modeling,kohl2012studio}.  

Here, we investigate the potential impact of these transformed approaches to laboratory instruction on students' epistemologies and expectations about the nature of \emph{experimental} physics.  To do this, we use the Colorado Learning Attitudes about Science Survey for Experimental Physics (E-CLASS) \cite{zwickl2014eclass,wilcox2016eclass}.  E-CLASS is a 30 item, Likert-style survey in which students are presented with a prompt (e.g., ``The primary purpose of doing physics experiments is to confirm previously known results.'') and asked to rate their level of agreement both from their personal perspective when doing experiment in class and that of a hypothetical experimental physicist.  The E-CLASS was developed in conjunction with laboratory course transformation efforts at the University of Colorado Boulder (CU) \cite{zwickl2013adlab}.  E-CLASS was validated through student interviews and expert review \cite{zwickl2014eclass}, and was tested for statistical validity and reliability using responses from students at multiple institutions and at multiple course levels \cite{wilcox2016eclass}.  This work is part of ongoing analysis of a growing, national data set of student responses to the E-CLASS.

\section{\label{sec:data}Data Sources \& Methods}

Data for this study were pulled from an existing data set of student responses to the E-CLASS collected using the E-CLASS online administration system \cite{wilcox2016admin} between 2013 and 2016.  In order to use this system, instructors completed a Course Information Survey (CIS) at the beginning of the course.  The CIS asked instructors to report logistical and demographic information about the course including instructional approaches used and course level (i.e., first year or beyond first year).  To explore how students' E-CLASS responses might vary across demographic lines, the system also asked students to provide demographic information about themselves including their major and gender.  While students were offered 15 distinct major options, we focus here on students' major as simply `physics' or `non-physics,' where physics includes both engineering physics and physics majors.  The primary motivation for this collapsing student majors into these binary categories was to preserve statistical power.  To account for the complex and non-binary nature of gender \cite{traxler2016gender}, students were provided three response options when reporting their gender: woman, man, and other.  Roughly 2\% ($N=87$) of the students selected the `other' option or left the question blank; thus, to preserve statistical power, the analysis with respect to gender will include only those students who selected `man' or `woman' when reporting their gender.  

\begin{table}[b]
\begin{ruledtabular}
\caption{Breakdown of students by major and gender in the matched first-year data set.  `Physics' includes both physics and engineering physics majors, and Non-physics includes other engineering, science, math, non-science, and undeclared.  A small number of students did not provide their major or gender; thus, the percentages will not sum to 100\%.  }\label{tab:major}
   \begin{tabular}{ l c  c c c c}
       & &\multicolumn{2}{c}{Major} & \multicolumn{2}{c}{Gender} \\
       & N &Physics& Non-Physics& Woman& Man \\
      \hline
      Transformed & 510 & 7\% & 91\% & 42\% & 56\% \\
      Traditional & 3091 & 8\% & 90\% & 42\% & 55\% \\
   \end{tabular}
\end{ruledtabular}
\end{table}

As the transformed instructional approaches described above have been designed primarily for introductory courses, we limited our analysis to the subset of E-CLASS data drawn from first-year labs covering introductory physics content.  Only students with matched pre- and post-instruction responses were included in the analysis.  The final, matched data set included $N=3601$ student responses from $60$ separate instances of $44$ distinct algebra- or calculus-based first-year courses spanning $29$ institutions.  Based on the instructors' responses on the CIS, each of these courses was classified as either Transformed or Traditional.  Courses in the Transformed category were those whose instructors reported using one or more of the transformed instructional approaches described above -- ISLE, Modeling, Studio Physics, or SCALE-UP.  Traditional courses were those whose instructors reported using only a traditional guided lab approach to instruction.  There were 20 Transformed and 40 Traditional courses in the data set; Table \ref{tab:major} shows the breakdown of students by major and gender across these courses.  

For scoring purposes, students' responses to each 5-point Likert item were condensed into a standardized, 3-point scale in which the responses `(dis)agree' and `strongly (dis)agree' were collapsed into a single category \cite{adams2006class}.  Students' responses to individual items were given a numerical score based on consistency with the accepted, expert response: favorable ($+1$), neutral ($+0$), or unfavorable ($-1$).  A student's overall score on the assessment is given by the sum of their scores on each of the 30 E-CLASS items resulting in a possible range of scores of $[-30,30]$.  For more information on the scoring of the E-CLASS see Ref. \cite{wilcox2016eclass}.  Because the distribution of scores on the E-CLASS is typically skewed towards positive scores, the following sections report statistical significance based on the non-parametric Mann-Whitney U test \cite{mann1947mwu} unless otherwise stated.  Where differences between means are statistically significant, we also report Cohen's $d$ \cite{cohen1988d} as a measure of effect size.

\section{\label{sec:results}Results}

The average overall pre- and post-instruction E-CLASS scores for Transformed and Traditional courses are given in Table \ref{tab:rawOverall}.  Students in Traditional courses had a statistically significant ($p\ll0.01$, $d=-0.3$), negative shift from pre- to post-instruction.  This negative shift over the course of one semester or quarter is consistent with historical trends for the E-CLASS \cite{wilcox2016eclass}.  Alternatively, students in Transformed courses did not shift significantly from pre- to post-instruction ($p=0.2$).  As shown in Table \ref{tab:rawOverall}, the difference between the pre-instruction scores for Traditional and Transformed was not statistically significant; however, the difference in post-instruction scores was significant with a moderate effect size.  

\begin{table}[b]
\begin{ruledtabular}
\caption{Raw pre- and post-instruction E-CLASS scores (points) for Transformed and Traditional courses.  Here, `Significance' refers to the statistical significance of the difference between the averages for students in Transformed and Traditional courses (Mann-Whitney U test).   }\label{tab:rawOverall}
   \begin{tabular}{ l c c c c }
       & Traditional & Transformed & Significance& Effect size \\
      \hline
      Pre & 16.4 & 17.1 & $p=0.07$ &   \\
      Post & 14.4 & 17.3 & $p\ll0.01$ & $d=0.4$  \\
   \end{tabular}
\end{ruledtabular}
\end{table}

\begin{figure*}%[b]
\includegraphics[width=\linewidth]{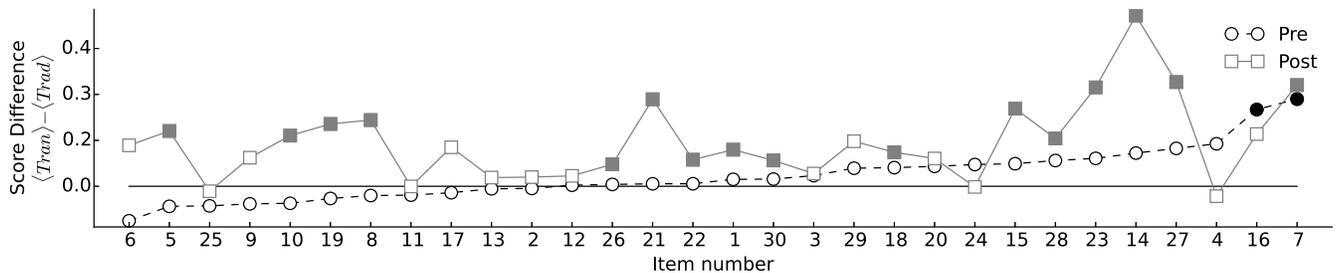}
\caption{Sorted plot of the difference between the average scores (points) of students in Transformed and Traditional courses for each item of the E-CLASS. Zero difference is marked by the solid horizontal line.  Filled markers indicate points for which the difference between the students in Traditional and Transformed courses is statistically significant (Mann-Whitney U \cite{mann1947mwu} and Holm-Bonferroni \cite{holm1979hb} corrected $p<0.05$).  See Ref.\ \cite{ECLASSwebsite} for full list of item prompts. }\label{fig:byItem}
\end{figure*}

We can also examine the difference between Transformed and Traditional courses on an item-by-item scale.  Fig.\ \ref{fig:byItem} shows the difference between the average scores of students in Transformed and Traditional course for each of the 30 items on the pre- and post-instruction E-CLASS.  Consistent with the results from the overall score, Fig.\ \ref{fig:byItem} shows that only two items show a statistically significant difference between the pre-instruction scores for students when split by instructional approach.  The magnitude of the difference in pre-instruction scores was small for both items (Cohen's $d=0.2$).  Alternatively, 16 of the 30 E-CLASS items show Transformed courses scoring significantly better at the end of the course (Mann-Whitney U \cite{mann1947mwu} and Holm-Bonferroni \cite{holm1979hb} corrected $p<0.05$).  The magnitude of the difference between post-instruction scores was moderate (Cohen's $d>0.3$) for only two of these items (Items 14 -- ``When doing an experiment, I usually think up my own questions to investigate'' and item 23 -- ``When I am doing an experiment, I try to make predictions to see if my results are reasonable'') and small for all other items ($d<0.3$).  Students in Traditional classes did not have significantly higher post-instruction means on any items.  

While comparison of the E-CLASS scores between Traditional and Transformed courses preliminarily indicates that Transformed courses achieved better gains overall and on the majority of items, previous work has shown that other factors, such as student major and gender, can also impact the comparison of overall E-CLASS means for certain subpopulations \cite{wilcox2016gender}.  To more clearly disentangle the relationship between instructional approach and post-instruction E-CLASS scores from that of other factors (e.g., pre-score, major), we used an analysis of covariance (ANCOVA) \cite{wildt1978ancova}.  ANCOVA is a statistical method for comparing the difference between population means while accounting for the variance associated with other variables.  Only students for whom we have matched E-CLASS scores along with both major and gender data were included in the ANCOVA ($N=3481$).  

\begin{table}[b]
\begin{ruledtabular}
\caption{Adjusted post-instruction E-CLASS means (points) from the 3-way ANCOVAs run on data from men and women separately.  Adjusted means were calculated while controlling for the variance associated with pre-instruction score and student major.  `Difference' represents the mean for students in Transformed courses minus the mean for students in Traditional courses. }\label{tab:gender}
   \begin{tabular}{ l c c c }
       & Traditional & Transformed & Difference \\
      \hline
      Women & 12.8 & 16.0 & 3.2 \\
      Men & 15.8 & 17.6 & 1.8 \\
   \end{tabular}
\end{ruledtabular}
\end{table}

We first ran a 4-way ANCOVA to compare post-instruction E-CLASS means for Traditional and Transformed courses while controlling for student major and gender, as well as pre-instruction score as a covariate.  To determine how these variables might also be related to one another, we included in the ANCOVA all possible interaction terms.  The 4-way ANCOVA revealed a statistically significant interaction between gender and instructional approach (F-test, $p=0.03$).  The presence of this interaction suggests that the relationship between instructional approach and post-instruction mean is different for men and women.  

To account for this interaction term, we split the data by gender and ran 3-way ANCOVAs for men and women separately.  Student major was a significant predictor of post-instruction scores for both men and women (F-test, $p=0.01$).  Instructional approach was also a significant predictor for both men and women (F-test, $p\ll0.01$), with a higher mean for students in Transformed courses in both cases.  However, the increase in post-instruction mean in Transformed courses was higher for women (see Table \ref{tab:gender}) and the difference was significant (as indicated by the statistical significance of the interaction between gender and instructional approach).  This finding supports the idea that both men and women benefited from Transformed instructional approaches, and this benefit was larger for women than men.  

Similarly, to determine the significance of gender as a predictor of post-instruction E-CLASS scores, we split the data by instructional approach and ran separate 3-way ANCOVAs for Transformed and Traditional courses separately.  For students in Traditional courses, adjusted post-instruction means for men were significantly higher than the adjusted means for women (F-test, $p<0.01$); however, for students in Transformed courses, post-instruction means for men and women were the same (F-test, $p=0.4$).  Thus, gender was a significant predictor of post-instruction E-CLASS means in Traditional courses only.

\section{\label{sec:discussion}Conclusions \& Limitation}

We analyzed student responses to a laboratory focused epistemologies assessment -- the E-CLASS -- to examine the relationship between the use of transformed instructional approaches and students' performance.  Data were drawn from a large-scale, national data set of students taking a first-year physics laboratory course.  We found that students in courses taught using transformed pedagogy had better post-instruction E-CLASS scores than students in courses taught using only traditional guide lab approaches to instruction.  This result was supported by the results of an ANCOVA, which examined the difference between post-instruction means while adjusting for the variance associated with pre-instruction scores, student major, and student gender.  The ANCOVA analysis showed that, when accounting for the variance associated with pre-instruction score and major, students in courses using transformed approaches to instruction had higher post-instruction means than those in courses using traditional approaches.  Moreover, the ANCOVA showed that this effect was larger for women than men suggesting that transformed pedagogies may have had a differentially positive effect on women.  Additionally, while gender was a significant predictor for Transformed courses, the differentially positive impact for women in Transformed courses made it so that gender was not a significant predictor for students in these courses.  

There are several limitations to the results reported here.  This study was limited to first-year (introductory) laboratory courses; additional data collection and analysis would be necessary to determine if these results hold in beyond-first-year courses.  Additionally, in order to preserve statistical power, all transformed pedagogies considered here (ISLE, Modeling Instruction, studio physics, and SCALE-UP) were aggregated together as a single group.  Thus, while it may be that one approach is more or less effective at promoting expert-like epistemologies and expectations, the current analysis cannot address this dynamic.  As data collection with the E-CLASS centralized administration system continues, comparisons between different transformed pedagogies may become possible. Finally, the purely quantitative analysis reported here cannot speak to how these transformed instructional approaches may have improved students' epistemologies relative to traditional approaches, nor can it determine why they differentially benefited women.  Determining the mechanism underlying the findings reported here will likely require additional research with a significant qualitative component.  

%\vspace{12pt}

\begin{acknowledgments}
This work was funded by the NSF-IUSE Grant DUE-1432204 and NSF Grant PHY-1125844.  Special thank you to the members of PER@C for all their feedback.  
\end{acknowledgments}

\bibliography{master-refs-ECLASS-5_16-mod}

\begin{thebibliography}{26}%
\makeatletter
\providecommand \@ifxundefined [1]{%
 \@ifx{#1\undefined}
}%
\providecommand \@ifnum [1]{%
 \ifnum #1\expandafter \@firstoftwo
 \else \expandafter \@secondoftwo
 \fi
}%
\providecommand \@ifx [1]{%
 \ifx #1\expandafter \@firstoftwo
 \else \expandafter \@secondoftwo
 \fi
}%
\providecommand \natexlab [1]{#1}%
\providecommand \enquote  [1]{``#1''}%
\providecommand \bibnamefont  [1]{#1}%
\providecommand \bibfnamefont [1]{#1}%
\providecommand \citenamefont [1]{#1}%
\providecommand \href@noop [0]{\@secondoftwo}%
\providecommand \href [0]{\begingroup \@sanitize@url \@href}%
\providecommand \@href[1]{\@@startlink{#1}\@@href}%
\providecommand \@@href[1]{\endgroup#1\@@endlink}%
\providecommand \@sanitize@url [0]{\catcode `\\12\catcode `\$12\catcode
  `\&12\catcode `\#12\catcode `\^12\catcode `\_12\catcode `\%12\relax}%
\providecommand \@@startlink[1]{}%
\providecommand \@@endlink[0]{}%
\providecommand \url  [0]{\begingroup\@sanitize@url \@url }%
\providecommand \@url [1]{\endgroup\@href {#1}{\urlprefix }}%
\providecommand \urlprefix  [0]{URL }%
\providecommand \Eprint [0]{\href }%
\providecommand \doibase [0]{http://dx.doi.org/}%
\providecommand \selectlanguage [0]{\@gobble}%
\providecommand \bibinfo  [0]{\@secondoftwo}%
\providecommand \bibfield  [0]{\@secondoftwo}%
\providecommand \translation [1]{[#1]}%
\providecommand \BibitemOpen [0]{}%
\providecommand \bibitemStop [0]{}%
\providecommand \bibitemNoStop [0]{.\EOS\space}%
\providecommand \EOS [0]{\spacefactor3000\relax}%
\providecommand \BibitemShut  [1]{\csname bibitem#1\endcsname}%
\let\auto@bib@innerbib\@empty
%</preamble>
\bibitem [{\citenamefont {Trumper}(2003)}]{trumper2003labs}%
  \BibitemOpen
  \bibfield  {author} {\bibinfo {author} {\bibfnamefont {R.}~\bibnamefont
  {Trumper}},\ }\href {\doibase 10.1023/A:1025692409001} {\bibfield  {journal}
  {\bibinfo  {journal} {Science {\&} Education}\ }\textbf {\bibinfo {volume}
  {12}},\ \bibinfo {pages} {645} (\bibinfo {year} {2003})}\BibitemShut
  {NoStop}%
\bibitem [{\citenamefont {{AAPT Committee on
  Laboratories}}(2015)}]{AAPT2015guidelines}%
  \BibitemOpen
  \bibfield  {author} {\bibinfo {author} {\bibnamefont {{AAPT Committee on
  Laboratories}}},\ }\href@noop {} {\enquote {\bibinfo {title} {{AAPT
  Recommendations for the Undergraduate Physics Laboratory Curriculum}},}\ }
  (\bibinfo {year} {2015})\BibitemShut {NoStop}%
\bibitem [{\citenamefont {Olson}\ and\ \citenamefont
  {Riordan}(2012)}]{olson2012excel}%
  \BibitemOpen
  \bibfield  {author} {\bibinfo {author} {\bibfnamefont {S.}~\bibnamefont
  {Olson}}\ and\ \bibinfo {author} {\bibfnamefont {D.~G.}\ \bibnamefont
  {Riordan}},\ }\href@noop {} {\bibfield  {journal} {\bibinfo  {journal}
  {Executive Office of the President}\ } (\bibinfo {year} {2012})}\BibitemShut
  {NoStop}%
\bibitem [{\citenamefont {{NRC Committee on Undergraduate Biology
  Education}}(2003)}]{nrc2003bio}%
  \BibitemOpen
  \bibfield  {author} {\bibinfo {author} {\bibnamefont {{NRC Committee on
  Undergraduate Biology Education}}},\ }\href@noop {} {\emph {\bibinfo {title}
  {BIO2010: Transforming undergraduate education for future research
  biologists}}}\ (\bibinfo  {publisher} {National Academies Press (US)},\
  \bibinfo {year} {2003})\BibitemShut {NoStop}%
\bibitem [{\citenamefont {Wieman}(2015)}]{wieman2015labs}%
  \BibitemOpen
  \bibfield  {author} {\bibinfo {author} {\bibfnamefont {C.}~\bibnamefont
  {Wieman}},\ }\href@noop {} {\bibfield  {journal} {\bibinfo  {journal} {The
  Physics Teacher}\ }\textbf {\bibinfo {volume} {53}} (\bibinfo {year}
  {2015})}\BibitemShut {NoStop}%
\bibitem [{\citenamefont {{NRC}}(2013)}]{nap2013nrc}%
  \BibitemOpen
  \bibfield  {author} {\bibinfo {author} {\bibnamefont {{NRC}}},\ }\href@noop
  {} {\emph {\bibinfo {title} {Adapting to a Changing World--Challenges and
  Opportunities in Undergraduate Physics Education}}}\ (\bibinfo  {publisher}
  {The National Academies Press},\ \bibinfo {year} {2013})\BibitemShut
  {NoStop}%
\bibitem [{\citenamefont {Etkina}\ and\ \citenamefont
  {Heuvelen}(2001)}]{etkina2001isle}%
  \BibitemOpen
  \bibfield  {author} {\bibinfo {author} {\bibfnamefont {E.}~\bibnamefont
  {Etkina}}\ and\ \bibinfo {author} {\bibfnamefont {A.~V.}\ \bibnamefont
  {Heuvelen}},\ }in\ \href@noop {} {\emph {\bibinfo {booktitle} {2001 PERC
  Proceedings}}}\ (\bibinfo {address} {Rochester, New York},\ \bibinfo {year}
  {2001})\BibitemShut {NoStop}%
\bibitem [{\citenamefont {Wells}\ \emph {et~al.}(1995)\citenamefont {Wells},
  \citenamefont {Hestenes},\ and\ \citenamefont
  {Swackhamer}}]{wells1995modeling}%
  \BibitemOpen
  \bibfield  {author} {\bibinfo {author} {\bibfnamefont {M.}~\bibnamefont
  {Wells}}, \bibinfo {author} {\bibfnamefont {D.}~\bibnamefont {Hestenes}}, \
  and\ \bibinfo {author} {\bibfnamefont {G.}~\bibnamefont {Swackhamer}},\
  }\href@noop {} {\bibfield  {journal} {\bibinfo  {journal} {AJP}\ }\textbf
  {\bibinfo {volume} {63}},\ \bibinfo {pages} {606} (\bibinfo {year}
  {1995})}\BibitemShut {NoStop}%
\bibitem [{\citenamefont {Wilson}(1994)}]{wilson1994studio}%
  \BibitemOpen
  \bibfield  {author} {\bibinfo {author} {\bibfnamefont {J.~M.}\ \bibnamefont
  {Wilson}},\ }\href {\doibase http://dx.doi.org/10.1119/1.2344100} {\bibfield
  {journal} {\bibinfo  {journal} {The Physics Teacher}\ }\textbf {\bibinfo
  {volume} {32}},\ \bibinfo {pages} {518} (\bibinfo {year} {1994})}\BibitemShut
  {NoStop}%
\bibitem [{\citenamefont {Beichner}\ \emph {et~al.}(2000)\citenamefont
  {Beichner}, \citenamefont {Saul}, \citenamefont {Allain}, \citenamefont
  {Deardorff},\ and\ \citenamefont {Abbott}}]{beichner2000scaleup}%
  \BibitemOpen
  \bibfield  {author} {\bibinfo {author} {\bibfnamefont {R.~J.}\ \bibnamefont
  {Beichner}}, \bibinfo {author} {\bibfnamefont {J.~M.}\ \bibnamefont {Saul}},
  \bibinfo {author} {\bibfnamefont {R.~J.}\ \bibnamefont {Allain}}, \bibinfo
  {author} {\bibfnamefont {D.~L.}\ \bibnamefont {Deardorff}}, \ and\ \bibinfo
  {author} {\bibfnamefont {D.~S.}\ \bibnamefont {Abbott}},\ }\href@noop {}
  {\bibfield  {journal} {\bibinfo  {journal} {2000 Proceedings of the ASEE}\ }
  (\bibinfo {year} {2000})}\BibitemShut {NoStop}%
\bibitem [{\citenamefont {Beichner}\ \emph {et~al.}(2007)\citenamefont
  {Beichner}, \citenamefont {Saul}, \citenamefont {Abbott}, \citenamefont
  {Morse}, \citenamefont {Deardorff}, \citenamefont {Allain}, \citenamefont
  {Bonham}, \citenamefont {Dancy},\ and\ \citenamefont
  {Risley}}]{beichner2007scaleup}%
  \BibitemOpen
  \bibfield  {author} {\bibinfo {author} {\bibfnamefont {R.~J.}\ \bibnamefont
  {Beichner}}, \bibinfo {author} {\bibfnamefont {J.~M.}\ \bibnamefont {Saul}},
  \bibinfo {author} {\bibfnamefont {D.~S.}\ \bibnamefont {Abbott}}, \bibinfo
  {author} {\bibfnamefont {J.~J.}\ \bibnamefont {Morse}}, \bibinfo {author}
  {\bibfnamefont {D.}~\bibnamefont {Deardorff}}, \bibinfo {author}
  {\bibfnamefont {R.~J.}\ \bibnamefont {Allain}}, \bibinfo {author}
  {\bibfnamefont {S.~W.}\ \bibnamefont {Bonham}}, \bibinfo {author}
  {\bibfnamefont {M.~H.}\ \bibnamefont {Dancy}}, \ and\ \bibinfo {author}
  {\bibfnamefont {J.~S.}\ \bibnamefont {Risley}},\ }\href@noop {} {\bibfield
  {journal} {\bibinfo  {journal} {Research-based reform of university physics}\
  }\textbf {\bibinfo {volume} {1}},\ \bibinfo {pages} {2} (\bibinfo {year}
  {2007})}\BibitemShut {NoStop}%
\bibitem [{\citenamefont {Adams}\ \emph {et~al.}(2006)\citenamefont {Adams},
  \citenamefont {Perkins}, \citenamefont {Podolefsky}, \citenamefont {Dubson},
  \citenamefont {Finkelstein},\ and\ \citenamefont {Wieman}}]{adams2006class}%
  \BibitemOpen
  \bibfield  {author} {\bibinfo {author} {\bibfnamefont {W.~K.}\ \bibnamefont
  {Adams}}, \bibinfo {author} {\bibfnamefont {K.~K.}\ \bibnamefont {Perkins}},
  \bibinfo {author} {\bibfnamefont {N.~S.}\ \bibnamefont {Podolefsky}},
  \bibinfo {author} {\bibfnamefont {M.}~\bibnamefont {Dubson}}, \bibinfo
  {author} {\bibfnamefont {N.~D.}\ \bibnamefont {Finkelstein}}, \ and\ \bibinfo
  {author} {\bibfnamefont {C.~E.}\ \bibnamefont {Wieman}},\ }\href@noop {}
  {\bibfield  {journal} {\bibinfo  {journal} {PRST-PER}\ }\textbf {\bibinfo
  {volume} {2}},\ \bibinfo {pages} {010101} (\bibinfo {year}
  {2006})}\BibitemShut {NoStop}%
\bibitem [{\citenamefont {Redish}\ \emph {et~al.}(1998)\citenamefont {Redish},
  \citenamefont {Saul},\ and\ \citenamefont {Steinberg}}]{redish1998mpex}%
  \BibitemOpen
  \bibfield  {author} {\bibinfo {author} {\bibfnamefont {E.~F.}\ \bibnamefont
  {Redish}}, \bibinfo {author} {\bibfnamefont {J.~M.}\ \bibnamefont {Saul}}, \
  and\ \bibinfo {author} {\bibfnamefont {R.~N.}\ \bibnamefont {Steinberg}},\
  }\href@noop {} {\bibfield  {journal} {\bibinfo  {journal} {American Journal
  of Physics}\ }\textbf {\bibinfo {volume} {66}},\ \bibinfo {pages} {212}
  (\bibinfo {year} {1998})}\BibitemShut {NoStop}%
\bibitem [{\citenamefont {Brewe}\ \emph {et~al.}(2009)\citenamefont {Brewe},
  \citenamefont {Kramer},\ and\ \citenamefont {O’Brien}}]{brewe2009modeling}%
  \BibitemOpen
  \bibfield  {author} {\bibinfo {author} {\bibfnamefont {E.}~\bibnamefont
  {Brewe}}, \bibinfo {author} {\bibfnamefont {L.}~\bibnamefont {Kramer}}, \
  and\ \bibinfo {author} {\bibfnamefont {G.}~\bibnamefont {O’Brien}},\
  }\href@noop {} {\bibfield  {journal} {\bibinfo  {journal} {PRST-PER}\
  }\textbf {\bibinfo {volume} {5}},\ \bibinfo {pages} {013102} (\bibinfo {year}
  {2009})}\BibitemShut {NoStop}%
\bibitem [{\citenamefont {Kohl}\ and\ \citenamefont
  {Vincent~Kuo}(2012)}]{kohl2012studio}%
  \BibitemOpen
  \bibfield  {author} {\bibinfo {author} {\bibfnamefont {P.~B.}\ \bibnamefont
  {Kohl}}\ and\ \bibinfo {author} {\bibfnamefont {H.}~\bibnamefont
  {Vincent~Kuo}},\ }\href@noop {} {\bibfield  {journal} {\bibinfo  {journal}
  {American Journal of Physics}\ }\textbf {\bibinfo {volume} {80}},\ \bibinfo
  {pages} {832} (\bibinfo {year} {2012})}\BibitemShut {NoStop}%
\bibitem [{\citenamefont {Zwickl}\ \emph {et~al.}(2014)\citenamefont {Zwickl},
  \citenamefont {Hirokawa}, \citenamefont {Finkelstein},\ and\ \citenamefont
  {Lewandowski}}]{zwickl2014eclass}%
  \BibitemOpen
  \bibfield  {author} {\bibinfo {author} {\bibfnamefont {B.~M.}\ \bibnamefont
  {Zwickl}}, \bibinfo {author} {\bibfnamefont {T.}~\bibnamefont {Hirokawa}},
  \bibinfo {author} {\bibfnamefont {N.}~\bibnamefont {Finkelstein}}, \ and\
  \bibinfo {author} {\bibfnamefont {H.}~\bibnamefont {Lewandowski}},\
  }\href@noop {} {\bibfield  {journal} {\bibinfo  {journal} {PRST-PER}\
  }\textbf {\bibinfo {volume} {10}},\ \bibinfo {pages} {010120} (\bibinfo
  {year} {2014})}\BibitemShut {NoStop}%
\bibitem [{\citenamefont {Wilcox}\ and\ \citenamefont
  {Lewandowski}(2016)}]{wilcox2016eclass}%
  \BibitemOpen
  \bibfield  {author} {\bibinfo {author} {\bibfnamefont {B.~R.}\ \bibnamefont
  {Wilcox}}\ and\ \bibinfo {author} {\bibfnamefont {H.~J.}\ \bibnamefont
  {Lewandowski}},\ }\href {\doibase 10.1103/PhysRevPhysEducRes.12.010123}
  {\bibfield  {journal} {\bibinfo  {journal} {PR-PER}\ }\textbf {\bibinfo
  {volume} {12}},\ \bibinfo {pages} {010123} (\bibinfo {year}
  {2016})}\BibitemShut {NoStop}%
\bibitem [{\citenamefont {Zwickl}\ \emph {et~al.}(2013)\citenamefont {Zwickl},
  \citenamefont {Finkelstein},\ and\ \citenamefont
  {Lewandowski}}]{zwickl2013adlab}%
  \BibitemOpen
  \bibfield  {author} {\bibinfo {author} {\bibfnamefont {B.~M.}\ \bibnamefont
  {Zwickl}}, \bibinfo {author} {\bibfnamefont {N.}~\bibnamefont {Finkelstein}},
  \ and\ \bibinfo {author} {\bibfnamefont {H.}~\bibnamefont {Lewandowski}},\
  }\href@noop {} {\bibfield  {journal} {\bibinfo  {journal} {AJP}\ }\textbf
  {\bibinfo {volume} {81}},\ \bibinfo {pages} {63} (\bibinfo {year}
  {2013})}\BibitemShut {NoStop}%
\bibitem [{\citenamefont {{Wilcox}}\ \emph {et~al.}(2016)\citenamefont
  {{Wilcox}}, \citenamefont {{Zwickl}}, \citenamefont {{Hobbs}}, \citenamefont
  {{Aiken}}, \citenamefont {{Welch}},\ and\ \citenamefont
  {{Lewandowski}}}]{wilcox2016admin}%
  \BibitemOpen
  \bibfield  {author} {\bibinfo {author} {\bibfnamefont {B.~R.}\ \bibnamefont
  {{Wilcox}}}, \bibinfo {author} {\bibfnamefont {B.~M.}\ \bibnamefont
  {{Zwickl}}}, \bibinfo {author} {\bibfnamefont {R.~D.}\ \bibnamefont
  {{Hobbs}}}, \bibinfo {author} {\bibfnamefont {J.~M.}\ \bibnamefont
  {{Aiken}}}, \bibinfo {author} {\bibfnamefont {N.~M.}\ \bibnamefont
  {{Welch}}}, \ and\ \bibinfo {author} {\bibfnamefont {H.~J.}\ \bibnamefont
  {{Lewandowski}}},\ }\href@noop {} {\bibfield  {journal} {\bibinfo  {journal}
  {ArXiv e-prints}\ } (\bibinfo {year} {2016})},\ \Eprint
  {http://arxiv.org/abs/1601.07896} {arXiv:1601.07896 [physics.ed-ph]}
  \BibitemShut {NoStop}%
\bibitem [{\citenamefont {{Traxler}}\ \emph {et~al.}(2015)\citenamefont
  {{Traxler}}, \citenamefont {{Cid}}, \citenamefont {{Blue}},\ and\
  \citenamefont {{Barthelemy}}}]{traxler2016gender}%
  \BibitemOpen
  \bibfield  {author} {\bibinfo {author} {\bibfnamefont {A.~L.}\ \bibnamefont
  {{Traxler}}}, \bibinfo {author} {\bibfnamefont {X.~C.}\ \bibnamefont
  {{Cid}}}, \bibinfo {author} {\bibfnamefont {J.}~\bibnamefont {{Blue}}}, \
  and\ \bibinfo {author} {\bibfnamefont {R.}~\bibnamefont {{Barthelemy}}},\
  }\href@noop {} {\bibfield  {journal} {\bibinfo  {journal} {ArXiv e-prints}\ }
  (\bibinfo {year} {2015})},\ \Eprint {http://arxiv.org/abs/1507.05107}
  {arXiv:1507.05107 [physics.ed-ph]} \BibitemShut {NoStop}%
\bibitem [{\citenamefont {Mann}\ and\ \citenamefont
  {Whitney}(1947)}]{mann1947mwu}%
  \BibitemOpen
  \bibfield  {author} {\bibinfo {author} {\bibfnamefont {H.~B.}\ \bibnamefont
  {Mann}}\ and\ \bibinfo {author} {\bibfnamefont {D.~R.}\ \bibnamefont
  {Whitney}},\ }\href@noop {} {\bibfield  {journal} {\bibinfo  {journal} {The
  annals of mathematical statistics}\ ,\ \bibinfo {pages} {50}} (\bibinfo
  {year} {1947})}\BibitemShut {NoStop}%
\bibitem [{\citenamefont {Cohen}(1988)}]{cohen1988d}%
  \BibitemOpen
  \bibfield  {author} {\bibinfo {author} {\bibfnamefont {J.}~\bibnamefont
  {Cohen}},\ }\href {https://books.google.com/books?id=Tl0N2lRAO9oC} {\emph
  {\bibinfo {title} {Statistical Power Analysis for the Behavioral Sciences}}}\
  (\bibinfo  {publisher} {L. Erlbaum Associates},\ \bibinfo {year}
  {1988})\BibitemShut {NoStop}%
\bibitem [{\citenamefont {Holm}(1979)}]{holm1979hb}%
  \BibitemOpen
  \bibfield  {author} {\bibinfo {author} {\bibfnamefont {S.}~\bibnamefont
  {Holm}},\ }\href@noop {} {\bibfield  {journal} {\bibinfo  {journal}
  {Scandinavian journal of statistics}\ ,\ \bibinfo {pages} {65}} (\bibinfo
  {year} {1979})}\BibitemShut {NoStop}%
\bibitem [{\citenamefont {{tinyurl.com/ECLASS-physics}}(2015)}]{ECLASSwebsite}%
  \BibitemOpen
  \bibfield  {author} {\bibinfo {author} {\bibnamefont
  {{tinyurl.com/ECLASS-physics}}},\ }\href@noop {} {} (\bibinfo {year}
  {2015})\BibitemShut {NoStop}%
\bibitem [{\citenamefont {{Wilcox}}\ and\ \citenamefont
  {{Lewandowski}}(2016)}]{wilcox2016gender}%
  \BibitemOpen
  \bibfield  {author} {\bibinfo {author} {\bibfnamefont {B.~R.}\ \bibnamefont
  {{Wilcox}}}\ and\ \bibinfo {author} {\bibfnamefont {H.~J.}\ \bibnamefont
  {{Lewandowski}}},\ }\href@noop {} {\bibfield  {journal} {\bibinfo  {journal}
  {ArXiv e-prints}\ } (\bibinfo {year} {2016})},\ \Eprint
  {http://arxiv.org/abs/1606.02629} {arXiv:1606.02629 [physics.ed-ph]}
  \BibitemShut {NoStop}%
\bibitem [{\citenamefont {Wildt}\ and\ \citenamefont
  {Ahtola}(1978)}]{wildt1978ancova}%
  \BibitemOpen
  \bibfield  {author} {\bibinfo {author} {\bibfnamefont {A.~R.}\ \bibnamefont
  {Wildt}}\ and\ \bibinfo {author} {\bibfnamefont {O.}~\bibnamefont {Ahtola}},\
  }\href@noop {} {\emph {\bibinfo {title} {Analysis of covariance}}},\
  Vol.~\bibinfo {volume} {12}\ (\bibinfo  {publisher} {Sage},\ \bibinfo {year}
  {1978})\BibitemShut {NoStop}%
\end{thebibliography}%

\end{document}